\documentclass[twocolumn,aps,prb,showpacs]{revtex4}

\usepackage{graphicx}
\usepackage{dcolumn}
\usepackage{bm}
\usepackage{color}
\input epsf
\input{epsf.sty}

\begin{document}

\preprint{APS/123-QED}

\title {Transfer of magnetization by spin injection between both
interfaces of a Ni nanowire}

\author{J.-E. Wegrowe, M. Dubey, T. Wade, H.-J. Drouhin, M. Konczykowski}
\affiliation{Laboratoire des Solides Irradi\'es, Ecole 
polytechnique, CNRS-UMR 7642 \& CEA/DSM/DRECAM, 91128 Palaiseau Cedex, France.}
\email{jean-eric.wegrowe@polytechnique.fr}

\date{\today} 

\begin{abstract}

 Magnetization switching provoked by spin-injection is studied in Ni
 nanowires of various size and morphology.  The response of the
 magnetization to the spin-injection is studied as a function of
 the amplitude of the current, the temperature, and the symmetry of the
 interfaces.  The amplitude of the response of the magnetization to
 spin-injection is a decreasing function of the temperature, does not
 depend on the current sign, and occurs only in the case of asymmetric 
 interfaces.  It is shown that the spin-injection does
 not act on small magnetic inhomogeneities inside the layer. Some 
 consequences in terms of longitudinal spin-transfer are discussed.
    
\end{abstract}


\maketitle

The possibility of driving magnetization states without the need of a
magnetic field was predicted some years ago by Berger \cite{Berger}
and independently by J. Slonczewski \cite{Sloncz}. This effect is observed 
today, especially with giant magnetoresistance (GMR) in nanopillar 
structures \cite{Albert}, and is interpreted 
in terms of {\it spin-transfer}. However,
current induced magnetization switching (CIMS) effects are also measured
in homogeneous $Ni$ and $Co$ nanowires  \cite{EPL,SPIE}, and in domain wall systems 
\cite{Klaui} where no GMR
can be measured.  Is the mechanism
responsible for CIMS different in these cases?  Previous studies tend
to show that both effects may be identical \cite{SPIE}.

The aim of this letter is to investigate further CIMS in a single
magnetic layer by measuring the amplitude of the effect as a function
of the temperature, of the amplitude and sign of the
injected current, of the symmetry of both interfaces,
and by studying the effect of the current on small magnetic
inhomogeneities inside the wire.  In order to modify the symmetry of
the interface with respect to the spin-injection, two kinds of
samples have been measured.  Samples of kind $A$ are $Au$(200 nm)/$Ni$(6
$\mu$m / $Ni$) and samples of kind B are $Au$(200 nm)/$Ni$(X $\mu$m)/$Cu$(6-X
$\mu$m / $Cu$) obtained by electrodeposition in nanoprous polycarbonate membrane
templates \cite{Meier}.  The membrane thickness is six micrometers, and the diameter
of the pores is about 60 nanometers.  The electrodeposited $Ni$ is
composed of small nanocrystallites, so that the system is quasi
amorphous with respect to the magnetic properties. The anisotropy is
then reduced to the shape anisotropy, defined by the aspect ration of
the nanowires \cite{PRL}. This is the reason 
why these experiments are performed with $Ni$, and not with $Co$
nanowires where CIMS are also observed. 

In samples of kind $A$, $Ni$ is electrodeposited up to the top of
the membrane, where a second thin $Au$ layer (45 nm) was deposited
previously.  The top $Au$ layer do not close the pores and its
function is to measure the potential between the top and the bottom of
the membrane during the electrodeposition, in order to obtained a
single wire contact.  The contact is formed by a $Ni$ mushroom of a
few hundreds of nm to one micrometer \cite{Travis}.  On the other
hand, in the sample of kind $B$ $Ni$ electrodeposition is stopped
after a calibrated time inside the pores, and the rest of the
deposition, including the contact, is performed with $Cu$.  Note
that $Au $ and the $Cu$ are identical with respect to the spin
injection (zero spin polarization of the current).  As a consequence,
in samples of kind $A$, the spin injection on the top interface is
performed with a geometry close to the current in plane of the
ferromagnetic layer (CIP), and with the geometry current perpendicular
to the plane (CPP) on the bottom interface.  In contrast, the samples
of kind $B$ have current injection with CPP geometry for both
interfaces, and are symmetric to that respect.

The hysteresis loop is measured with the anisotropic
magnetoresistance with a field applied at about $\theta = 82^{O} \pm 4 $
from the wire axis (Fig.  1).  This angle is chosen because the signal
due to the irreversible magnetization reversal reaches its maximum
amplitude \cite{PRL}. The resistance of the sample is 148 $\Omega$ at 
150K, in agreement with the resistivity of $Ni$ and the wire diameter 
of 60 nanometers.  Both half-hysteresis loops (for decreasing
fields and for increasing fields) are symmetric, and are composed by a
reversible part (equilibrium states of the magnetization), and a
single irreversible jump, which occurs at the switching field
$H_{sw}$.  The sample is chosen in such a way that all equilibrium
states are uniform magnetization states within about two percent of
the total magnetization.  This can be checked by fitting the envelope
of the curve with the uniform reversal model, and applying the curve
obtained for different angles of the external field. The irreversible jump
occurs at a critical angle $\varphi_{c}$, related to a critical field
$H_{sw}(\theta)$ (spinodal limit).  Detailed studies of the magnetic
states are reported elsewhere \cite{PRL}.   The magnetic
configurations are uniform for all temperatures, but the anisotropy of the
$Ni$ nanowire increases dramatically when the temperature is decreased
from room temperature down to Helium temperature, as observed in 
previous studies \cite{Meier}. Beyond the change in magnetic anisotropy, the AMR 
decreases from  1.5 \% at 200 K down to 1\% at 4.5K in contrast to 
the measured bulk AMR \cite{Potter} :
this is due to the  {\it finite size effect} as predicted and quantified
in reference \cite{Rijks}.

The effect of the current is studied by injecting a pulsed current of
one microsecond duration (about 100 ns rising time) at a given
magnetization state.  The magnetization state is described 
equivalently by its
angle \cite{EPL} or by the
distance $\Delta H $ to the switching field (Fig.  1).  This distance
is a measure of the barrier hight to be
overcome by the magnetization switch \cite{SPIE}, due to current injection,
from one equilibrium state to the other (two states system).  Consequently, the maximum
distance $\Delta H_{max} $ where the magnetization reversal can still
be observed gives the amplitude of the effect of the current on the
magnetization. 

\begin{figure}[h]
\centerline{\epsfxsize 8cm \epsfbox{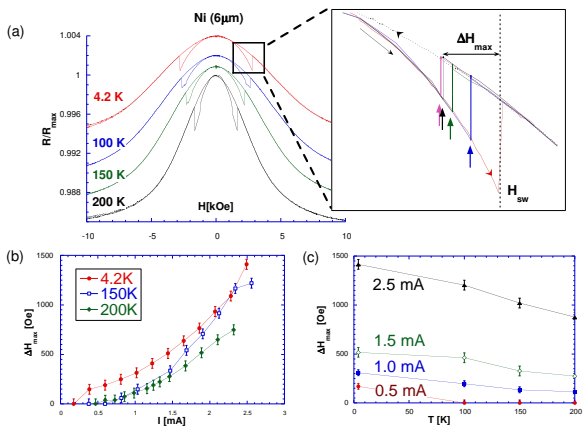}}
\caption{(a) Hysteresis loop at various temperature of the 6 $\mu$ m 
$Ni$ nanowire (the curves are shifted for clarity).  The experimental
protocol is illustrated in the inset (magnification of the 
irreversible part of the hysteresis).  The arrows indicate the field
at which a current pulse is injected.  The effect of the current
injection is to switch the magnetization if the field is located in
the region $\Delta H_{max}$. (b) $\Delta H_{max}$ as a function of the 
current amplitude. (c) $\Delta H_{max}$ as a function of the 
temperature}
\label{}
\end{figure}

The parameter $\Delta H_{max} $ is plotted as a function of the
current amplitude (Fig.  1(b)) for different temperatures.  The curves
are reproducible, and { \it do not depend on the sign of the current}! 
The amplitude of the effect is of the order of 1 kOe (0.1 Tesla) for 
2 mA injection, while the maximum induced field (Oersted field 
produced radialy by the current) is 
below 100 Oe (0.01 Tesla) and has a radial symmetry. The effect of the induced 
field as been ruled-out in previous works \cite{EPL,SPIE}.
The slope $\Delta H_{max}/ \Delta I $is about 500 Oe/ mA (or about 25
mT/(10$^{-7}$ A/cm$^{2}$)) which is typical in such samples
\cite{EPL}, and is { \it of the same order of magnitude as what has
been measured in pillar structures} ( in the same units : 20, 20, 16
reported in \cite{Albert}).  The temperature
dependence (Fig.  1(c)) varies from a factor 1.5 between 200 K and
4.2K. This temperature dependence has not been reported so far in
nanowires, and is opposite to that measured in pillar samples
\cite{Albert}.  This observation seems to indicate that CIMS is
directly related to the spin diffusion length.

\begin{figure}[h]
\centerline{\epsfxsize 8cm \epsfbox{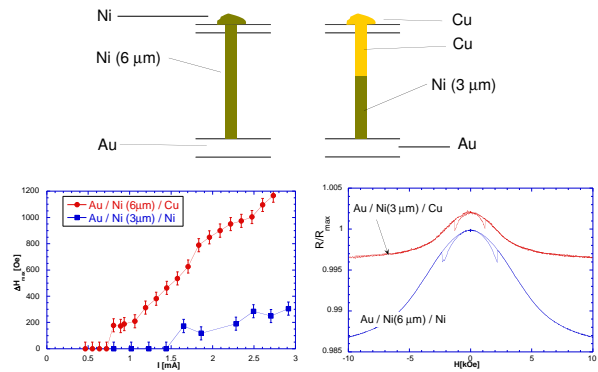}}
\caption{(a) Comparison of the effect of current injection in sample of 
type A : $Au$ / $Ni$(6 $\mu $m) / $Ni$) and type B : $Au$ / $Ni$(3 $\mu $m) / $Cu$ 
and (b) comparison of the hysteresis loops of the sample of type A and 
of type B}
\end{figure}

The comparison between samples of kind $A$ and samples of kind $B$
(resistance R=112 $\Omega$ at 150 K) shows that there is a qualitative
difference (Fig 2(a)) of the response of the magnetization to the
current excitation.  The hysteresis loops show the variation in terms
of AMR (the reduction in sample $B$ is due to the fact that the active
part of the device is reduced by a factor 2, with approximately
constant total resistance), and in terms of anisotropy due to the
reduction of the aspect ratio of the $Ni$ nanowire.  A statistical
ensemble of samples of both kinds have been measured, with varying the
parameters of the deposition, like the pH and the concentration of the
solution.  A series of samples of kind $B$ have been measured with
varying the size of the $Ni$ layer.  The result is always similar to
that presented in Fig.  2(a).  The response of the magnetization to
the current injection is negligible in the case of the samples of kind
$B$, and this result is { \it independent of the anisotropy, and more
generally independent of the energy barrier of the $Ni$ magnetic layer}
(proportional to the volume of the layer).  In all cases, the curves
are reproducible, and do not depend on the current direction.

\begin{figure}[h]
\centerline{\epsfxsize 8cm \epsfbox{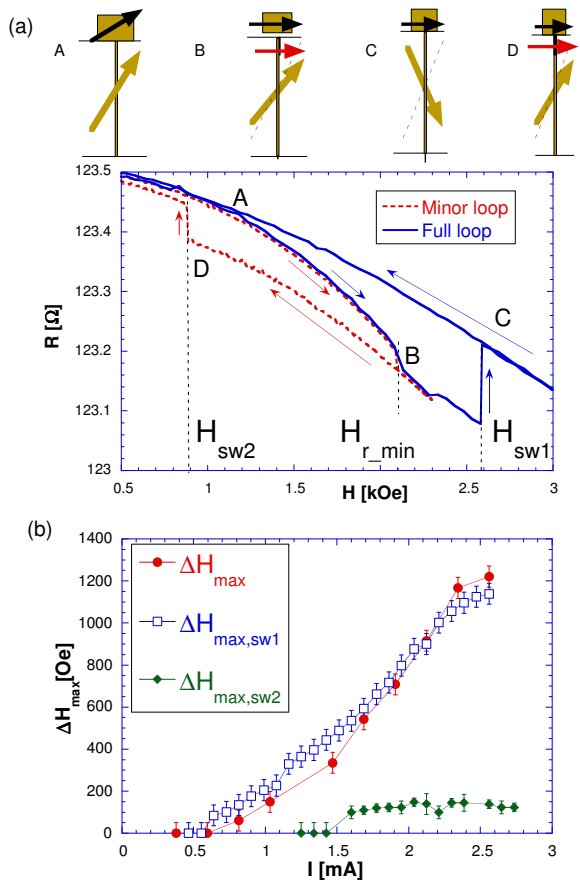}}
\caption{Effect of the current injection on a small inhomogeneity due 
to the $Ni$ contact. The nucleation and annihilation of the inhomogeity is evidenced by its 
minor loop.  }
\end{figure}

In order to investigate further the role of the $Ni$ contact on the
top of the membrane, we focalized our attention to the presence of a
small inhomogeneity in the wire (Fig.  3).  The envelop of the
hysteresis loop is identical to the usual ones, but a zoom of the
irreversible part of the hysteresis shows the nucleation of an
inhomogeneity, which should be a pinned vortex or any constrained
domain wall in the top of the membrane (the position of the
inhomegeneity has not been directly measured, but $Ni$ wires with Cu
contacts do not exhibit such structures).  The deviation to the
uniform rotation of the magnetization shows that the inhomogeneity
inside the wire is of the order of 2 \% of the total magnetization, i. 
e. about 100 nm (the magnetization of the $Ni$ contact is not measured
since the current density is negligible).  A minor loop can be
obtained if the external field is swept back after the creation of the
inhomogeneity (field $H_{r-min}$), but before its annihilation at
higher field.  The minor hysteresis loop describes then the different
magnetization states with the inhomogeneity (sketched in Fig.  3(a)). 
(A) the magnetization of the $Ni$ contact makes an angle close to that
of the $Ni$ nanowire, there is no significant inhomogeneity.  (B) The
magnetization of the $Ni$ contact is in the direction of the applied
field.  Exchange coupling induced an inhomogenoeity inside the wire
(point B).  The annihilation of the inhomogeneity occurs at higher
fields because the angle of the magnetization of the nanowire is close
to the external field.  If the external field is swept back before the
annihilation field, the inhomogeneity is maintained (and compressed)
from point B to point D, at which the homogeneous state is restored.

The response of the whole magnetization to the current injection is not 
significantly affected by the inhomogeneity. If $\Delta H_{max,sw1}$ 
is the amplitude of the effect after following the minor loop (i.e. 
with the presence of the inhomogeneity) and $\Delta H_{max}$ is the 
direct loop (no inhomogeneity in the region AB), we have $ \Delta 
H_{max,sw1} \approx  \Delta H_{max}$ (Fig. 3(b)). Furthermore, the current 
injection does not act on the annihilation (D) of the inhomogeneity : the 
jump $ H_{sw2}$ is not significantly affected by the current 
injection (and the variation $\Delta H_{max,sw2} \approx 180$  Oe can be 
attributed to the field induced by the current).

In conclusion, we have to deal with the three following observations :
(1) the sign of the current does not play any role (in contrast to
CIMS in GMR pillar structures), (2) the presence of the asymmetry is
necessary for CIMS, (3) the current acts on the whole structure
composed by the $Ni$ wire and both interfaces, and not locally.  All
occurs as if CIMS effects were due to an imbalance of the
spin-injection between both interfaces of the $ Ni$ layer.  In this
picture, a spin transfer occurs at the first normal/ferromagnet
interface, which should be compensated at the second
ferromagnetic/normal interface.  If an imbalance exists between both
interfaces due to an asymmetry with respect to spin injection, a {\it
longitudinal spin transfer} can be expected from the current to the
magnetic layer.  This hypothesis is corroborated by the fact that in
the GMR pillars (in contrast to homogeneous nanowires)
the change in the direction of the current leads to a change in the
response of the magnetization as much as changing the
magnetic configuration from parallel to antiparallel \cite{SPIE}. 
Indeed, as described by Berger \cite{Beger}, the asymmetry between the two interfaces (in terms of
spin-polarization) is not only due to the spin injection $\tilde {\Delta
\mu} $ \cite{Berger,wegrowe} but also to the spin accumulation $
\Delta \mu $ where the sign depends on the sign of the current.  It is
then expected that, in contrast to the homogeneous wires, the imbalance
due to the asymmetry is reversed by changing the sign of the current.


\begin{references}
    
\bibitem{Berger} L. Berger, J. Appl. Phys.,  {\bf 55}, 1954 (1984),  L. 
Berger, Phys. Rev. B {\bf 54}, 9353 (1996) and L. Berger J. Appl. 
Phys. {\bf 93}, 7693 (2003).
	
	
\bibitem{Sloncz} J. C. Slonczewski, J. Magn. Magn. Mat. {\bf 159} L1 (1996).

\bibitem{Albert} F. J. Albert, J. A. Katine, R. A. Buhrman, D. C. 
Ralph, Appl. Phys. Lett. {\bf 77} 3809 (2000), J. Grollier, V. Cros, A. Hamzic, J.M. George, H. 
Jaffes, A. Fert, G. Faini, J. Ben Youssef, H. Le Gall, Appl. Phys. 
Lett. {\bf 78}, 3663 (2001), J. Z. Sun, D. J. Monsma, D. W. Abraham, M. J. Rooks and 
R. H. Koch, Appl. Phys. Lett. {\bf 81}, 2202 (2002), J.-E. Wegrowe, A. Fabian, X. Hoffer, Ph.  Guittienne, D. Kelly, E. Olive, J.-Ph.  Ansermet
Appl. Phys. Lett. {\bf 91}, 6806 (2002), S. Urazhdin, H. Kurt, W. P. Pratt, and J. 
Bass, Appl. Phys. Lett. {\bf 83} 114 (2003), S. Urazhdin et al. 
cond-mat/03031492.

\bibitem{EPL} J-E Wegrowe, D. Kelly, Y. Jaccard, Ph. Guittienne, J-Ph Ansermet, 
Europhys. Lett. {\bf 45}, 626 (1999), J-E Wegrowe et al. Europhys.  Lett.  
{\bf 56}, 748 (2001) and D. Kelly et al. accepted for publication in 
PRB.
	
\bibitem{SPIE} J.-E. Wegrowe cond-mat/0306103.

\bibitem{Klaui} M. Kla\"ui et al. Appl. phys. Lett. {\bf 83} (2003), 
105,  J. Grollier et al. Appl. Phys. Lett. {\bf 83} (2003). 

\bibitem{Meier} J. Meier, B. Doudin, and J.-Ph. Ansermet, J. Appl. Phys. 79, 6010 (1996).

\bibitem{PRL} J.-E. Wegrowe, D. Kelly, A. Franck, S.E. Gilbert, J.-Ph. Ansermet, 
	 Phys. Rev. Lett. {\bf 82} 3681(1999) and Y. Jaccard et al. Phys. 
	 Rev B 62, 1141 (2000).
	 	 
\bibitem{Travis} Ch. Sch\"onenberger et al., J. Phys. Chem. B, 101 (1997), 5497.

 	
\bibitem{Potter} T. R. McGuire and R. I. Potter, IEEE Trans. 
Mag.-11 (1975), 1018.

\bibitem{Rijks} Th. G. S. M. Rijks, R. Coehoorn, M. J. M. de Jong, W. 
J. M. de Jonge, Phys. Rev. B, 51 (1995), 283.

\bibitem{wegrowe} J.-E. Wegrowe, Phys. Rev. B {\bf 62}, 1067 (2000).


\end{references}
\end{document}